\documentclass[prc,12pt,superscriptaddress,onecolumn,showpacs,amsmath,amssymb,aps,nofootinbib,floatfix]{revtex4-2}
\usepackage{tikz}
\usepackage{appendix}
\usepackage{epsfig}
\usepackage{xcolor}
\usepackage{amsmath}
\usepackage{mathtools,slashed}
\usepackage{amssymb}
\newcommand{\bra}[1]{  \left| #1 \right> }
\newcommand{\ket}[1]{  \left< #1 \right| }
\newcommand{\be}{\begin{eqnarray}}
 \newcommand{\secref}[1]{Section\,(\ref{#1})}
 \newcommand{\eqn}[1]{Eq.\,(\ref{#1})}

\newcommand{\ensemble}[1]{\left\{ #1 \right\}}

\newcommand{\order}[1]{ \mathcal{O} \left( #1 \right) }
\newcommand{\lnz}{\ln \mathcal{Z}}

\newcommand{\ee}{\end{eqnarray}}

\newcommand{\ave}[1]{\left\langle #1 \right\rangle}
\newcommand{\abs}[1]{\left| #1 \right|}
\newcommand{\eqcomma}{\phantom{AA},\phantom{AA}}

 \usepackage[colorlinks=true, linkcolor=blue, citecolor=blue, urlcolor=blue]{hyperref}
\begin{document}

\title{Gaussian fluctuating Generally covariant diffusion}
\author{Giorgio Torrieri,David Montenegro}
\affiliation{Universidade Estadual de Campinas - Instituto de Física Gleb Wataghin\\
Rua Sérgio Buarque de Holanda, 777\\
 CEP 13083-859 - Campinas - São Paulo - Brasil}

%\date{\today}
%\date{April 2025}
\begin{abstract}
We extend the previously developed \cite{sampaio} generally covariant formalism to include diffusion of conserved charges.  We construct a partition function with chemical potential, and calculate the dynamics of the diffusion matrix in terms of the Ward identity enforcing charge conservation and linear response.  We comment on the differences between diffusion of conserved scalar charges and full viscous hydrodynamics, and outline possible physical applications
\end{abstract}
%\date{November 2025}

\maketitle
\section{The problem of relativistic diffusion \label{intro}}
Relativistic diffusion is of considerable phenomenological and theoretical interest, being the leading non-equilibrium correction in transport of conserved charges in both heavy ion collisions and neutron stars \cite{expdiff1,expdiff2,expdiff3,expdiff4}.   It is also an important standalone aspect of the theoretical study of non-equilibrium relativistic statistical mechanics. 

It is often said \cite{forster,heinz,kovtun} that diffusion is the simplest hydrodynamic-like theory, because it is governed by a conservation law equation, combined with thermodynamics.  However, diffusion has no flow but just,as the name says, diffusive motion,    A conserved current $J_\nu$`s evolution is defined by
\begin{equation}
\label{diffu}
\partial_\nu J^\nu=0 \eqcomma 
\vec{J_i} = -D\vec{\nabla} \mu_Q
\end{equation}
 where $\mu$ is the chemical potential, a Lagrange multiplier enforcing charge conservation of the equilibrium distribution.   Just like in hydrodynamics, conservation laws and an expression derivable from statistical mechanics, namely the dependence of the diffusion coefficient $D$ from the chemical potential $\mu_Q$, is enough to close the equations and make the system solvable from arbitrary initial conditions.
 The form of $D$ is usually thought to arise from a Kubo type formula
 \begin{equation}
 \label{kubod}
D = \lim_{k \rightarrow 0} \frac{1}{k} \mathrm{Im} \int e^{ikx} \ave{\left[J_x(x,t),J_x(0,0)\right]}d^4 x
\end{equation}
In the full local thermalization limit $D\rightarrow 0$ no diffusion occurs (physically ``many`` microscopic collisions stop the particle).

  This theory is much simpler than hydrodynamics (where $J_\mu = T \rho \beta_\mu+\order{\partial}...$ and $\rho$ is the density while $\beta_\mu =u_\mu/T$ is the flow divided by temperature), as, because of the absence of flow, it`s leading order form is linear, and solvable exactly with Greens functions techniques.
  
It also has the usual issues of theories based on thermalization have with Lorentz invariance and causality\cite{gavathermal}, and,because of its simplicity, is a good laboratory to overcome them.   A usual way to address them is \cite{gava,grozdiff,xinan,delacretaz,lubl,kovtundiff} via a relaxational dynamics of the Maxwell-Cattaneo type
\begin{equation}\label{iscur}
\tau_Q \frac{d \vec{J}}{dt}+\vec{J}=\left. \vec{J}\right|_{diffusion}=-D\vec{\nabla}\mu
\end{equation}
which restores causality for a long enough $\tau_Q$ w.r.t. $\sqrt{D}$ at the price of promoting $J_\nu$ to an independent degree of freedom w.r.t. $\mu_Q$, albeit one with relaxational dynamics only (if hydrodynamics is included, $J_\perp$, the perpendicular component of $J$ w.r.t. flow, haa the role of such a degree of freedom).  However the Lorentz invariance issue never really goes away, since there is a ``special`` frame in which the chemical potential is defined.   At least in the microcanonical limit, the underlyng dynamics should however be Lorentz invariant (and in the canonical and grand canonical limit dynamics should be in terms of the {\em relative} velocity between the system and the bath).

In \cite{ergodic,dore,sampaio,crooks} a way to restore Lorentz invariance was proposed in the context of fluid dynamics:   Since ideal fluid behavior is equivalent to local ergodicity in any foliation \cite{ergodic}, one should be able to restore Lorentz invariance by assuming any dissipative dynamics is locally indistinguishable from a fluctuation.  Thus, if one treats fluctuations and average values on the same footing, one can impose Lorentz invariance by imposing such a local equivalence relation.     In this work we shall apply this framework to the simplified problem of diffusion.  In \secref{problem} we qualitatively dscribe the relationship between non-perturbative fluctuations and general covariance, in \secref{solution} we show that the issues described above can b clarified in this approach.  Finally in \secref{disc} we discuss how our calculation fits in the results of \cite{sampaio} as well as other approaches of relativistic statistical mechanics.   Finally, in the appendix we make a detailed comparison between the approach presented in this work and the more usual expansion around a thermostatic background on which gradient and Schwinger-Keldysh effective theories are based on
\section{Fluctuations and general covariance\label{problem}}
The physical interpretation of this general covariance is that ultimately the ``microscopic physics`` is ergodic for a perfectly thermalized system, but, because of the ``andromeda paradox`` \cite{andromeda} such ergodicity is frame dependent.   Only by promoting Lorentz invariance to invariance under all possible foliations does a Lorentz-invariant definition of local equilibrium emerge\cite{ergodic}.  This definition is microcanonical, but can be generalized to the canonical \cite{sampaio} ensemble provided fluctuations in the relative velocity between the system and the bath are accounted for in a Lorentz-invariant way.  For dynamics to be generally covariant, it has to be governed by the Ward identity.  In \cite{sampaio} it was shown that this identity, together with linear response, was enough to specify the dynamical evolution of the system provided the partition function defined at every cell has a Gaussian form.
 Note that this does not mean that 3-point functions and higher are not there, but just that they are exclusively functions of 2-point functions, as shown quantitatively in \cite{xinan}.

 A simple justification of the Gaussian ansatz is that, since general covariance requires fluctuations, this ansatz is the simplest that can explicitly accomodate it.
Furthermore,  Gaussian partition functions seem to work well even in strongly coupled dynamics \cite{gauss1} and are also the only  finite width
distribution which is a fixed point of renormalization group flows \cite{gauss2} (a result similar to Donsker's theorem in mathematics) we believe it to be a reasonable ansatz in the  absence of critical points.   
What this means physically is that taking the lattice spacing on which to define the Gaussian partition functions to zero will not result in additional terms but in the ``flow`` of the average and width to a specific fixed point \cite{gauss2}. 
It is worth noting that, as shown in the appendix, existing thermostatic backreaction approaches \cite{kovtun,delacretaz,xinan} are also Gaussian in the sense that ''non-Gaussian terms'' either regularize the two leading cumulants (as in stochastic diffusion), or just the first cumulant (when an effective deterministic theory is written out as a gradient expansion, reflecting the spectral analysis of the Kubo correlator).
Where our approach differs is that fluctuations are non-perturbative, and the general covariance underlying the local Lorentz invariance of the ''ergodic'' microscopic dynamics is exact.   This makes for a much more complicated dynamics, where the whole ensemble needs to be evolved from initial conditions (In \cite{kovtun,delacretaz,xinan} it is directly computable from the action), but avoids a thermostatic background thereby respecting the microscopic symmetries of the dynamics.
Note also that going beyond the Gaussian approximation is possible, provided higher order Ward identities and response relations are available.

Here we want to apply the methods of \cite{sampaio} to diffusion of conserved charges.   
We would like to immediately clarify that in a sense this is an academic exercise:  The physical applicability would be a system whose energy-momentum fluctuation and dissipation scales are very short, but conserved charge fluctuation and dissipation scales are very long.  We are not aware of such a system in nature, for instance in the cited nuclear and astrophysic application \cite{expdiff1,expdiff2,expdiff3,expdiff4} diffusion of conserved charge is of the same order as diffusion of momentum.

Nevertheless, the importance of diffusion as the simplest hydrodynamic-like theory motivates us to see how it looks like in this new ansatz, and also to see how the semplification of diffusion w.r.t. full hydrodynamics reflects here.  As we shall see, indeed the simpler structure of diffusion means the dynamics will have some qualitative differences.
\section{Gaussian fluctuating generally covariant diffusion\label{solution}}
We consider $J_\mu$ to be an operator, endowed with a partition function, whose conserved charge part is  
\begin{equation}
\label{zubpart}
\mathcal{Z}= \mathrm{Tr} \exp \left[ -\int_\Sigma d\Sigma_\nu \left( \mu_Q \hat{J}^\nu\right) \right],
\end{equation}
(where the trace denotes a functional integral over all microscopic field configurations)

moreover, we postulate that this partition function is approximately.gaussian
\begin{equation}
\label{gaussansatz}
\mathcal{Z}_J\simeq \exp\left[ -\frac{(\left( J_\mu - \ave{J_\mu (x,t)}\right)(\left( J_\nu - \ave{J_\nu(x,t)}\right) }{D^{\mu \nu}(x,t)} \right]
\end{equation}
Thus we obtain equations of motion for $\ave{J_\mu (x,t)}, D^{\mu \nu}(x,t)$ from the following requirements
\begin{description}
\item[General covariance]  $\ave{J_\mu (x,t)}, D^{\mu \nu}(x,t)$ should transform covariantly under general space-time refoliations
\item[Fluctuation/dissipation relation] representing thermal equilibrium. Locally, one cannot tell whether a deviation from equilibrium is an ergodic fluctuation or dissipative dynamics
\item[Charge conservation] for {\em every} element of the ensemble.   This will be enforced by a Ward identity
\end{description}
The first requirement comes from the definition of the global partition function using the Zubarev formalism \eqn{zubpart} must hold for arbitrary cell choices $d^3 \Sigma_\mu$.

if $\Sigma_\mu$ is a space-like 4-vector \cite{ergodic,crooks}
\begin{equation}
\Sigma_\mu=(t(x),\vec{x})=(t(\phi_I),\phi_{I=1,2,3}),
\end{equation}
(in its own rest frame $\Sigma_\mu=(0,x,y,z)$) its exterior derivative (in its own rest-frame $d\Sigma_\mu=(dV,0,0,0)$) is time-like.  
which also implies a general metric,
\begin{equation}
\label{foldef}
g_{\mu \nu}= \frac{\partial \Sigma_\mu}{\partial \Sigma^\nu}   \eqcomma d^{3} \Sigma_\mu = d\phi_I d\phi_J d\phi_K \epsilon_{\mu \nu \alpha \beta}\epsilon_{IJK}  
\frac{\partial \Sigma^\nu}{\partial \phi_I}
\frac{\partial \Sigma^\alpha}{\partial \phi_J}
\frac{\partial \Sigma^\beta}{\partial \phi_K},
\end{equation}
It then follows that, since $g_{\mu \nu}=\partial \Sigma_\mu/\partial \Sigma_\nu$, that gneral covariancee automatically implies invariance under $d^3 \Sigma_\mu$.

The linear response equation, written in a covariant form, is therefore
\begin{equation}
\label{jlin}
\hat{J}_\mu(\Sigma_0) = \ave{J_\mu(\Sigma_0)}+\int d\Sigma'_0 D_{\mu \nu}(\Sigma,\Sigma') \delta \hat{A}^\nu (\Sigma'_0) d\Sigma'_0\eqcomma D_{\mu \nu}(\Sigma,\Sigma`)= \ave{\left[J_\mu (\Sigma_0) J_\nu (\Sigma'_0)\right]}
\end{equation}
note that if $\Sigma_0$ is a space-like direction, one can regard this as an auto-correlation, if it is in a time-like direction, it will be a linear response evolution subject to causality.  The two situations can be incorporated in the same object by analytically continuing $D_{\mu \nu}(\Sigma,\Sigma') $ from $\Sigma_0$ to $\Sigma_0(1+i\epsilon)$ in the time-like direction, as in \cite{forster}.

The LHS of \eqn{jlin} has a hat, because we have an ensemble of configurations of $J_\mu$ at every point of spacetime.   We can therefore think of \eqn{jlin} as a representation of a ``typical`` ensemble element.  It will be the average,plus a random term represented by a random excitation of an auxiliary Gauge field, repressented by $\delta \hat{A}^\nu (\Sigma'_0)$.
But now we have to remember that conservation of charge is {\em exact}, true for {\em every} element of the ensemble but not just the average.

We therefore contract \eqn{jlin} with a volume element, and enforce charge conservation
\begin{equation}
\label{jconf}
\frac{d}{d\Sigma_0} \left[ \hat{J}_\mu(\Sigma_0) d^3 \Sigma^\mu\right] =0 \eqcomma \frac{d}{d\Sigma_0} Q=0
\end{equation}
This means that there is a locally conserved quantity $\delta \mathcal{Q}(x)$, defined and conserved at the ensemble level, which is also a scalar under general refoliations
\begin{equation}
\label{jconf2}
\delta \mathcal{Q}= \ave{J_\mu(\Sigma_0)}d^3 \Sigma^\mu+  d^3\Sigma^\mu \int d\Sigma'_0  
D_{\mu \nu}(\Sigma,\Sigma')
 \delta \hat{A}^\nu (\Sigma'_0) 
\end{equation}
note that locally $d^3 \Sigma_\mu d\Sigma_0$ is a scalar, so $d^3 \Sigma_\mu d\Sigma_0'$ reflects ``tidal`` effects of metric deformations, it would be a second derivative.   However, $\delta \hat{A}^\nu$ is simply an auxiliary field (In case of hydrodynamics\cite{sampaio} it is a physical metric perturbation )

Physically, Eq. \ref{jconf2} means that in an arbitrarily defined ``fluid cell`` one might observe momentary violations of charge, which could be statistical fluctuations (you did not know what the charge was in the first place!) or diffusive exchanges with the rest of the system. Changing the cell definition (i.e. the foliation) changes the preferred interpretation, but it should have no effect on the dynamics (this interpretation ambiguity is at the heart of why the "Andromeda paradox" ambiguity \cite{andromeda} does not affect dynamics. It would if the system was purely deterministic).

So, $\mathcal{Q}$ is the constraint between the average and fluctuation of the ensemble that enforces $Q$ conservation for each element of the ensemble.   For instance, the classic linear response result eq. 2.5 of \cite{kovtun} is reproduced for $\delta \mathcal{Q}=0$ and a thermal background, $d^3\Sigma_\mu=(d^3 x,\vec{0})$.

Eq. \eqn{jlin} together with \eqn{jconf2},together with a Gaussian ansatz
\label{gaussianansatz} (which will mean $\ave{\left[J_\mu (\Sigma_0) J_\nu (\Sigma'_0)\right]}$ is a function of $D_{\mu \nu}(x,t)$) will allow $\ave{J_\mu}(x,t),D_{\mu\nu}(x,t)$ to be integrated from an initial condition  $\ave{J_\mu}(x,t_0),D_{\mu\nu}(x,t_0)$.  The dynamics will be foliation independent, although of course the directions of $\ave{J_\mu}(x,t_0),D_{\mu\nu}(x,t_0)$ will transform covariantly with the foliation.

To see how it works, we note that $\delta A^\mu$ is an auxiliary field, whose energy density must be positive.   Without loss of generality, we therefore introduce a foluation metric\footnote{Note that for the energy-momentum tensor this trick is impossible, because it is a rank 2 tensor, necessitating volume preserving diffeomorphisms \cite{sampaio}.  On the other hand, in that case the gravitational Ward identity acts as a further constraint, as seen between eqs 52 and 53 of \cite{heinz}}
\begin{equation}
d^3 \Sigma'_\mu =d^3 x \frac{\delta A_\mu}{\sqrt{A_\alpha A^\alpha}} \eqcomma g_{\mu \nu}=\frac{\partial \Sigma'_\mu}{\partial \Sigma^\nu}
\end{equation}
\eqn{jconf2} then becomes independent of auxiliary gauge fields
\begin{equation}
\label{jconf3}
\frac{d^3 \mathcal{Q}}{d^3 \Sigma_\mu}= \ave{J_\mu(\Sigma_0)}+  \int d\Sigma'_0 d^3 \Sigma'_\mu \mathcal{D}_{\mu \nu}(\Sigma,\Sigma')   =0
\end{equation}
$\mathcal{D}_{\mu \nu}$ can be obtained from $D_{\mu \nu}$ by the usual procedure \cite{forster,sampaio} ($\tilde{f}$ is the Fourier transform of $f$)
\begin{equation}
\mathcal{\tilde{D}}_{\mu \nu}=\frac{1}{2i} \left( 
\frac{\tilde{D}^{  \mu \nu}(\Sigma_0,k)}{\tilde{D}^{  \mu \nu}(-i\epsilon\Sigma_0,k)}-1\right)
\end{equation}
Since $d\Sigma_0 d\Sigma_1 d\Sigma_2 d\Sigma_3$ is a Lorentz scalar, one can see that the integration measure is a Lorentz scalar times a unit 4-vector, ensuring the right transformation properties.

Thus the system of equations can be closed without the need for volume-preserving diffeomorphisms (as in \cite{sampaio}).   Also, in a sense the Ward identity and the fluctuation-dissipation relation needed for the propagation of the ensemble of $J$ coincide, a feature of the much simpler structure of $J_{\mu}$ w.r.t. $T_{\mu \nu}$.  Because flow and co-moving frames do not appear, the violation of general covariance by the integration measure pointed out in \cite{sampaio} (and argued heuristically there to be small) does not appear here.
\section{Discussion \label{disc}}
We have achieved the aim set forward in section \ref{problem}
Given an ensemble $\ensemble{J_{\mu}}$, characterized by an initial value of $\ave{J_\mu(x,t_0)},D_{\mu \nu}(x,t_0)$ \eqn{jconf3} together with an analytical continuation can be evolved (i.e. $\ave{J_\mu(x,t)},D_{\mu \nu}(x,t)$ can be obtained for arbitrary $t>t_0$)  in a generally covariant way (i.e. the evolution of the ensemble as a whole will not depend on the foliation chosen).   For such a calculation, however, the thermal spectral function of $\ave{\left[\tilde{J}_\mu(k),\tilde{J}_\nu(k')\right]}$ is needed to {\em arbitratry order} (the thermal average can then be transformed to arbitrary foliations using the tensor technique, see Fig. 2 and 3 of \cite{sampaio}).  This is a good illustration of the fact that the cumulant and gradient expansion do not commute.  

A numerical implementation of this model, just as of \cite{sampaio}, is therefore well beyond the scope of this work.
As shown in \cite{sampaio}, however, such a dynamics has a simple physical interpretation in terms of Crooks fluctuation theorem \cite{crooks}.
Given a proper time $d\tau^2 \propto d^3 oigma_\mu d^3 \Sigma^\mu$ (the constant of proportionality is a scale of the dimension of a 4-volume, hence a Lorentz scalar), the probability of a configuration going forward $P\left(\ensemble{\mu_Q},\tau+d\tau|\ensemble{\mu'_Q},\tau\right)$ rather than in reverse $P\left(\ensemble{\mu_Q}',\tau+d\tau|\ensemble{\mu_Q},\tau\right)$ is the exponential of the relative entropy, which in the case of diffusion alone is related trivially to the chemical potential via the susceptibility $\chi_Q$
\begin{equation}
\frac{P\left(\ensemble{\mu_Q},\tau+d\tau|\ensemble{\mu_Q}',\tau\right)}{P\left(\ensemble{\mu_Q}',\tau+d\tau|\ensemble{\mu_Q},\tau\right)}=\exp[\left[S(\ensemble{\mu_Q}')-S(\ensemble{\mu_Q}\right]
\end{equation}
where the entropy of a field of $\ensemble{\mu_Q}$ defined locally in $\Sigma$ is defined in terms of
the volume hypersurface element $d^3 \Sigma_\mu$ and it`s normal unit vector $n_\mu$
\begin{equation}
 S(\ensemble{\mu_Q})=\int d^3\Sigma_\mu n^\mu \frac{1}{T(\Sigma)}\int_0^{\rho_Q(\Sigma)} \rho_Q(T,\mu'_Q(\Sigma))d\mu'_Q
\end{equation}
(In hydrodynamics $\beta_\mu$ would take the place of $n_\mu$).

In the full local equilibration limit, $D_{\mu \nu}\rightarrow 0$ and $\ave{J}$. is conserved exactly,as expected.  Note that in this limit both fluctuations and deviations from equilibrium vanish togehter, as expected from the general covariance requirement.
Close to local equilibrium, for small fluctuations phenomena such as long-time tails should appear \cite{grozdiff}.
However, in parallel to \cite{sampaio},in a strongly coupled regime with non-negligible fluctuations long-time tails might become {\em less} likely, since fluctuations of the mean background foliation can not be neglected.  Such fluctuations would need to be subtracted from any long-time tail signal and, analgously to ghosts in gauge theory correlators, could weaken the ``physically observable`` tails discussed in  \cite{grozdiff}.

One can also view Eq. \eqn{jconf3} as a Wiener process
\begin{equation}
\label{wiener}
X_t = \mu t+ \sigma \hat{W}_t
\end{equation}
with the role of $X_t =Q_\tau \rightarrow J_\mu d^{3} \Sigma_\mu$ the drift $\mu \rightarrow \ave{J_\mu} d^{3} \Sigma_\mu$ and the stochastic variation $\Sigma \rightarrow D_{\mu \nu} d^{3} \Sigma^\mu d^{3} \Sigma'_\nu$.
In such a picture, the long-time evolution of $\ave{J_\mu}$ must be, because of general covariance, independent of the exact slicing of $\Sigma_\mu$.   It is clear that this requirement is exactly equivalent to the property of hyperbolicity for the partial differential equation governing $\ave{J_\mu}$.
Thus the requirement of strong hyperbolicity  \cite{geroch,disconzi} is more general than kinetic theory embedding \cite{gavhyp} and is inextricably linked with the Lorentz invariance of the underlying dynamics.
We note that the non-Relativistic average only diffusion equation, much like the non-relativistic Navier-Stokes equation lacks this property, as expected since there absolute time and space have quite different roles.   It is thus not surprising that historic development of diffusion and hydrodynamic equations found it so difficoult to accomodate local Lorentz invariance, which nevertheless must be there since microscopic properties that lead to approximate ergodicity are Lorentz invariant.  The missing ingredient was that fluctuations are an integral part of the dynamics.

Note that hydrodynamic flow makes no appearance in these equations.   This is because hydrodynamic flow (or more correctly $\beta_\mu$) can be thought-of as Lagrange multiplier for 4-momentum current.   But here we are only concerned with a conserved scalar charge.  In this limit, because there is no flow, there is also no freedom to redefine the flow frame, which is the basis for the general covariance symmetry \cite{dore} on which this work is based.  However, the discussion here makes it clear that, as long as the conversed charge is a Lorentz scalar and fluctuations are included, general covariance is realized without considering the flow frame,just with the foliation.

However the absence of flow also implies that a notion of local equilibrium in a purely diffusion theory is not possible. This is not surprising:  As shown in \cite{gavathermal} the zeroth law of thermodynamics, on which the notion of equilibrium is based, can be defined relativistically via the existence of a killing vector, which is equivalent to a hydrodynamic flow velocity. 
\cite{gavathermal} also notes that this killing vector can be defined only if the two volume cells exchange energy momentum vis uncharged mediators,while diffusion presupposes a Conserved charge.

Away from the purely diffusive limit, in case both flow and diffusion are present, and the partition function has the form 
\begin{align}\label{genpfunc}
\mathcal{Z} &=\mathrm{Tr} \exp\left[ -d^3\Sigma_\nu \left( \beta_\mu T^{\mu \nu}(\phi) + \mu_Q J^\mu (\phi) \right)\right] \simeq \notag \\ 
& \simeq   \exp\left[ - C_{\mu \nu,\alpha,\beta}\left(\Sigma,\Sigma'\right) \left(T^{\mu \nu}(\Sigma)-\ave{T^{\mu \nu}(\Sigma')}\right)  \left(T^{\alpha \beta}(\Sigma')-\ave{T^{\alpha\beta}(\Sigma')}\right)\right]\times \notag \\ &\times \exp \left[ - D_{\mu \nu} \left( J^{\mu} (\Sigma)-\ave{J^\mu(\Sigma)}\right) \left( J^\nu (\Sigma')-
\ave{J^{\nu}(\Sigma'}\right) \right] 
\end{align}
The calculations goes much as before, but then the fluctuation tensor $D_{\alpha \beta}$ will be in the
frame at rest with the flow, just like the object $C'_{\alpha \beta \mu \gamma}$ on which dynamics depends in \cite{sampaio}.  

The energy-momentum and conserved current term decouple but, because this frame is not in general the same as where $D_{\alpha \beta}$ is diagonal, we will have \cite{sampaio} 
\begin{equation}
\label{gengausssolved}
\lnz(\Sigma)=\int_\Sigma d\lnz(\Sigma) \eqcomma \textcolor{black}{d \mathcal{Z} (d\Sigma)}=  
\left( \sqrt{ \prod_i \lambda_i^2 }\sqrt{ C_{\alpha \beta}^{'\alpha \beta} }\right)^{-1} \times
\left( \sqrt{ \prod_i \ave{J_i}^2 \times \mathrm{det}_{\alpha \beta}\abs{D_{\alpha \beta} }}\right)^{-1}
\end{equation}
where $C'$ , as in \cite{sampaio}, is the correlator in the rest frame w.r.t. $\beta_\mu$ and $\lambda_i$ are the eigenvalues (related to the energy density and pressure).  In the Wiener picture \eqn{wiener} we need two correlated 4-vectors $d³ \Sigma_\mu$ and $\beta_\mu$ to maintain Lorentz covariance.  However, as long as these vectors are chosen appropriately (as argued in \cite{sampaio}) the evolution will be independent of the choide of $\Sigma_\mu$.   This is a demonstration that $\beta_\mu$ can be somewhat arbitrary as long as long term averages in fluctuating hydrodynamics must obey hyperbolic partial differential equations, something argued from the purely deterministic theory \cite{geroch,disconzi}.  Note that in the non-relativistic limit hyperbolicity is not necessary since there is a preferred foliation (that going forward in the absolute time) and indeed non-relativistic hydrodynamics is non-hyperbolic.

As in \cite{sampaio} the language seems very different from usual hydrodynamics (the appendix provides a matching) ,but it is in fact representing the same thing, with the addition that fluctuations are included non-perturbatively.In particular the component
of $D_{\mu \nu}$ perpendicular to velocity is the one associated with the diffusion constant in the usual Kubo formulae.
\begin{equation}
D= \lim_{k,w \rightarrow 0} \frac{1}{k} \mathrm{Im} \int e^{iwt-kx} D_\perp \left( \Sigma=(t,x),\Sigma'(0,0)\right) \eqcomma \beta_\mu D_\perp^{\mu \nu} =0
\end{equation}
However, in the full theory fluctuations in energy and conserved charge will intertwine in a way depending on the equation of state, leading to a complicated correlation pattern between fluid cells not readily captured by such asymptotic expansions.   In analogy to \cite{sampaio}, it is reasonable to suppose that in the strongly coupled but fluctuating limit $D_\perp$ is smaller than the Kubo valu
 because the flow fluctuations and diffusion fluctuations cancel out to some extent when the physical J,T correlators are calculated  (this is a similar argument to the one used above to say that long-time tails are suppressed).

The discussion above makes it apparent why no equivalent of the pseudo-gauge symmetry exists for 1-currents (this issue was discussed in \cite{jeonspin}).  A redefinition $J_\mu
\rightarrow J_\mu+\partial_\nu B^{\mu \nu}$ (as made in \cite{jeonspin}), where $B^{\mu \nu}$ is a fully antisymmetric tensor, would be
irrelevant since $\Gamma_{\mu \alpha \beta} B^{\mu \nu}=0$; 
 Ultimately this reflects the metricity condition in Riemmanian geometry, so changing $\Sigma_\mu$ would produce no effect
on the dynamics. This is of course not true if the energy-momentum tensor is thought to have an anti-symmetric component,
representing a near-equilibrium spin densityThe analysis of this situation has been conducted elsewhere \cite{spinwork}.

The extension of the Gaussian approach to chemical potential opens up the perspective to study systems with phase transitions and critical points \cite{critpoint}.   For this, however, one needs to explicitly go beyond Gaussian dynamics \cite{critpoint2}, and consider parititon functions containing non-trivial Skewness and Kurtosis.   This is in principle possible along a more involved procedure of the one outlined here, using 4th order Ward identities and quadratic response theories (4 parameters, corresponding to the 4 cumulants, 2nd and 4th order Ward identities and linear and quadratic transport coefficients should close the stochastic equaions).   However, in this case the partition function can not be integrated in closed form, and the evolution equations are likely to be coupled integro-differential equations.    This is a much more involved problem, left for future work.

In conclusion, we extended the formalism of \cite{sampaio} to include diffusion of conserved charge quantities.    This reinforces the conclusions made in that work, namely that for full covariance average quantities and fluctuations must be considered on an equal bases, and for idal behavior the two must go to zero together respecting conservation laws.  As a result, it is not automatic that systems with more degrees of freedom are also the closest to local equilibrium.

We are in theprocess to extend this formalism further, encompassing spin \cite{spinwork}, multiple conserved currents with non-trivial symmetry groups, and gauge symmetries \cite{ghosts},and eventually to a non-trivial phase diagram.

This work was was initiated while on a visit to the School of Nuclear Science and Technology, Lanzhou University, as an answer to questions from Navid Abbasi. it would have been impossible without Navid`s hospitality and fruitfuil,insightful and deep discussions.  GT thanks
Bolsa de produtividade CNPQ 305731/2023-8 and FAPESP 2023/06278-2as well as participation in the tematico  2023/13749-1 for support.
\appendix

\section{Stochastic model for linearized hydrodynamics in a hydrostatic background}

\subsection{Diffusion Correlation function without noise}

The linear response equation, written in a covariant form, is

 \begin{align}\label{12345}
D_{\mu\nu} (\Sigma,\Sigma^\prime)  &= -i \Theta (\Sigma^\prime_0 - \Sigma_0) G_{\mu\nu}  (\Sigma,\Sigma^\prime) =  \ket{} J^\mu (\Sigma )  , j^\nu ( \Sigma^\prime)   \bra{}  = \int \frac{d k^4}{(2 \pi)^4} \chi^{\prime\prime}_{\mu\nu} e^{ik \textbf{r} - i \omega t}   
\end{align}
where $\chi^{\prime\prime}_{\mu\nu}$ is the absorptive part. It is convenient to work with the full complex dynamic susceptibility $ \chi_{\mu\nu} = \chi^\prime_{\mu\nu} + i \chi^{\prime\prime}_{\mu\nu} $. We employ the Kramers-Kronig relation to evaluate the real and imaginary part 
\begin{equation}\label{21}
\chi_{\mu\nu} (k, \omega) =  - \mathcal{P} \int \frac{d \omega^\prime}{\pi} \frac{\chi^{\prime\prime}_{\mu\nu} (k, \omega^\prime) }{\omega^\prime - z } \eqcomma \chi^{\prime\prime}_{\mu\nu} (k, \omega) =  - \mathcal{P} \int \frac{d \omega^\prime}{\pi} \frac{\chi^\prime_{\mu\nu} (k, \omega^\prime) }{\omega^\prime - \omega }
\end{equation}
The equation \eqref{jlin} becomes  

\begin{equation}
\hat{J}_\mu(\Sigma_0) = \ave{J_\mu(\Sigma_0)}\hat{I} + \int d\Sigma^\prime_0 \int \frac{d \omega}{(2 \pi)}  \int \frac{d^3 \textbf{k}}{(2 \pi)^3} \chi^{\prime\prime}_{\mu\nu} e^{ik \textbf{r} - i \omega \Sigma^\prime_0} \delta \hat{A}^\nu (\Sigma^\prime_0)    
\end{equation}
Where $\hat{I}$ is the unit operator.
Before applying the Laplace transformation, we perform the transformation indicated below

\begin{equation}
d \Sigma_0 =  \hat{e}_0  \times \  d t, \eqcomma  \delta \hat{A}_\nu (\Sigma^\prime_0) = \int \frac{d \omega^\prime}{(2 \pi)} e^{i \omega^\prime \Sigma^\prime_0} \delta  \hat{A}_\nu (\omega^\prime)
\end{equation}
In this way, we find
\begin{align}\label{123}
\hat{J}_\mu(\Sigma_0) &= \ave{J_\mu(\Sigma_0)}\hat{I} + \int d t  \int \frac{d \omega}{(2 \pi)}  \int \frac{d^3 \textbf{k}}{(2 \pi)^3} \chi^{\prime\prime}_{\mu\nu} e^{ik \textbf{r} - i \omega t} \int \frac{d \omega^\prime}{(2 \pi)}  e^{i \omega^\prime t} \delta \hat{A}^\nu (\omega^\prime) \notag    \\
 &= \ave{J_\mu(\Sigma_0)} \hat{I}+ \hat{e}_0 \int \frac{d \omega}{2 \pi i}  \frac{\chi^{\prime\prime}_{\mu\nu} (k,\omega)}{\omega(\omega - z)} \delta \hat{A}^\nu ( \omega)  
\end{align}
Assuming $\chi_{\mu\nu} $ as a smooth function of $\omega$, we can apply the identity 
\begin{equation}
    \lim_{ \epsilon \rightarrow{0}} \frac{1}{\omega^\prime - \omega - i \epsilon} = \mathcal{P} \frac{1}{\omega^\prime - \omega} + i \pi \delta (\omega^\prime - \omega )
\end{equation}
where $\mathcal{P}$ is the principal value since the poles $z$ are above the real axis
\begin{equation}\label{ref}
Re \Big[ \frac{\delta \hat{J}^\mu (\textbf{k},\omega + i \epsilon)}{\delta \hat{A}^\nu (\omega)} \Big] = \frac{\chi^{\prime\prime}_{\mu\nu}}{\omega}
\end{equation}
Rewriting \eqref{ref} as an expansion for large values of $z$, we get the analytical result  

\begin{align}
& \lim_{z \to \infty} \frac{ \delta \hat{J}^\mu (k,z)}{\delta \hat{A}_\nu (z)} = \notag \\
&  \frac{i}{z} \chi_{\mu\nu} (k) + \frac{i}{z^2} \int \frac{d \omega}{\pi} \chi^{\prime\prime}_{\mu\nu} (k,\omega) +  \frac{i}{z^2} \int \omega \frac{d \omega}{\pi} \chi^{\prime\prime}_{\mu\nu} (k,\omega) +  \frac{i}{z^3} \int \frac{\omega^2}{(z-\omega)} \frac{d \omega}{\pi} \chi^{\prime\prime}_{\mu\nu}  (k,\omega) + \ldots 
\end{align}
Following this idea, it is straightforward to determine the dynamical susceptibility 
\begin{align}
\chi^{\prime\prime}_{\mu\nu} (k, \omega) =& \frac{D}{\xi^2} + \frac{\chi_{\mu\nu}  D k^2 }{\omega^2 + D^2 [ k^2  - (\chi \sqrt{(p^2/p^2_0)}/ \langle n \rangle \xi^2 )   ]} + \frac{\chi_{\mu\nu} \omega }{\omega^2 - (D/\xi^2 )\sqrt{J^\mu (x,0) J_\mu (x,0) }}  \notag \\ 
& + \frac{ \chi_{\mu\nu} \omega^2 \xi^2}{\omega^3 + i \omega^2 ( D^2 k^2 + \xi^2) - \omega (D k^2 - (J^\mu (x,0) J_\mu (x,0))^{3/2} ) }
\end{align}
Which matches standard derivations, for example \cite{forster}.
\subsection{Correlation function with relaxation time}

To avoid violation of stability and causality , we assume diffusive forces to not turn on instantaneusly and include the relaxation time for a gradient to become a current. By enforcing charge conservation in \eqref{jconf}, we can represent \eqref{iscur} in a convenient form  
 
\begin{equation}
 d^3 \mathcal{Q} \bigg|^{\infty}_{- \infty} = \int d^3 \Sigma_\mu \frac{d^3 \mathcal{Q}}{d^3 \Sigma_\mu} = \int d \Sigma_0 \ave{J_\mu(\Sigma_0)} +  \int d^3 \Sigma^\prime_\mu \int d\Sigma^\prime_0  \mathcal{D}_{\mu \nu}(\Sigma,\Sigma^\prime) 
\end{equation}
which yields
\begin{equation}  
\int d^3 \Sigma_\mu    \int d\Sigma_0 \ket{} J^\mu (0) , J^\nu (\Sigma)  \bra{} = \int d\Sigma_0 \int d^3 \Sigma_\mu   \bigg[   \frac{\partial^2}{\partial \Sigma_0^2 }  - \frac{1}{\tau_Q} \bigg( \frac{\partial}{\partial \Sigma_0} - D \nabla^2 \bigg)    \bigg] \ave{\hat{J}^\mu (\Sigma_0)}  = 0
\end{equation}
Applying the Laplace transform and integrating by parts, we obtain
\begin{align}
%0 = & \frac{\delta A_\nu (t)}{\sqrt{A_\alpha A^\alpha}}  \bigg[ \int d^3 x e^{- i k \textbf{x}} \int d t e^{it z} \bigg( -iz + D k^2 - \tau z^2 \bigg) \hat{J}^\mu (x, t)  - \int d^3 x e^{- i  k \textbf{x}} \hat{J}^\mu (x, 0)    \bigg]  \notag \\ & + \hat{J}^\mu(\Sigma_0) \int d^3 x e^{- i k \textbf{x}}  \int d t e^{it z} \bigg( -iz - \tau z^2 \bigg) \frac{\delta A_\nu (t)}{\sqrt{A_\alpha A^\alpha}} \notag \\ 
 0 = & \delta A_\nu (\Sigma_0)  \bigg[ \bigg( -iz + D k^2 - \tau z^2 \bigg) \hat{J}^\mu (k, z)  -    \hat{J}^\mu (x, 0)    \bigg] + \hat{J}^\mu(\Sigma_0) \bigg( -iz - \tau z^2 \bigg) \delta A_\nu (z)
\end{align}
We write the current in function of the source 

\begin{equation}\label{label21}
    \frac{\hat{J}^\mu (k,z)}{\delta A_\nu (z)} =  \frac{ \chi_{\mu\nu} (i z + z^2 \tau ) }{(D k^2 - i z - \tau z^2 ) - (\tau/z) \sqrt{J^\mu (x,0) J_\mu (x,0) }  } 
\end{equation}
According to \eqref{ref}, the dynamic susceptibility of \eqref{label21} can be determined by performing an expansion for large values of $z$
\begin{equation}
\frac{\hat{J}^\mu (k,z)}{\delta A_\nu (z)} = \bigg[ \frac{i}{z} + \frac{i D k^2}{\tau z^3} + \frac{i }{\tau z^2} \sqrt{J^\mu (x,0) J_\mu (x,0) } 
+ \frac{i D \tau  }{\tau z^3} (J^\mu (x,0) J_\mu (x,0))^{3/2}  + \mathcal{O} \bigg( \frac{1}{z^4} \bigg) \bigg]
\end{equation}
which matches similar theories developed for example in \cite{gava}.

\subsection{Correlation function with noise}

We introduce random currents in equilibrium, so that the linearized constitutive relation for the current takes the form when $\tau_{Q} \ne 0$
\begin{equation}
\vec{J}(\omega, {\bf k}) = \frac{ -D\vec{\nabla} \nu + \partial_k \vec{r}^k }{1-i \tau_Q \omega} {\rm .} 
\end{equation}
where $\vec{r}^k$ is the Gaussian stochastic vector in space-time with the ensemble average $\langle \vec{r}^k  \rangle = 0$. Since we can separate the noise from the low energy effective action, the effective action rises by integrating out the fast modes  
\begin{align}\label{pqppqp}
& e^{-S_{\mathrm{eff}}[x_{\mathrm{slow}}]} =
\int \mathcal{D}x_{\mathrm{fast}} \;
e^{-S[x_{\mathrm{slow}} + x_{\mathrm{fast}}]} \\
& S[x_{\mathrm{slow}} + x_{\mathrm{fast}}]
= S[x_{\mathrm{slow}}]
+ \frac{1}{2} \int dt\,dt'\; x_{\mathrm{fast}}(t)\,
\mathcal{O}(t,t')\, x_{\mathrm{fast}}(t')
+ \mathcal{O}(x_{\mathrm{fast}}^3).
\end{align}
The generating functional now becomes
\begin{equation}\label{cfr}
Z = \int\! D\xi\, D\phi \, D \Bar{\phi} ; e^{ - S_{eff} [\phi, \Bar{\phi}] }\; e^{-S_\xi[r^k]}
\end{equation}
where $\phi=(v_i,\delta T,\mu)$. Assuming the initial conditions are uncorrelated. Since the noise is a Gaussian functional integral, its probability distribution is
\begin{equation}\label{noisexi}
 \langle \dots \rangle = \int
  D \xi \; e^{-S_\xi[r^k]} \dots, \quad   S_\xi[r] = \frac12 \int\!\! dt\, d^dx \left(
  \frac{r_i \bar\sigma_{ij}^{-1} r_j}{2\bar T}
   \right)    
\end{equation}  
where $\bar\sigma_{ij}$ controls the strength of stochastic fluctuations. The $rr$ two-point function is
\begin{equation}
\langle r_i (x,t)\, r_k (x',t') \rangle =  \mathcal{G} \delta(x - x^\prime) \delta( t - t^\prime) 
\end{equation} 
where $\mathcal{G}$  determines the strength of the noise. The representation of the correlation function in \eqref{cfr} after the integration of the noise \eqref{noisexi} is
\begin{equation}\label{efaction}
G_{J J}(\Sigma, \Sigma^\prime) = \langle J(\Sigma)\, J( \Sigma^\prime ) \rangle = \int D\phi\,D \Bar{\phi}
  \langle \delta({\rm e.o.m.}) \rangle\, |J| \, e^{-S_{\rm eff}[\phi,\Bar{\phi}]}
  J(\Sigma  )\, J( \Sigma^\prime )  
\end{equation}
 and the stress
tensor and the current are given by the classical linear constitutive relations. Splitting the velocity perturbations into $\delta u = \delta u_L + \delta u_T$, we have 
$\delta u_L^i(\omega, {\bf k}) \equiv {\bf k} \cdot {\bf \delta u}/|{\bf k}|$ and $\delta u_T(\omega, {\bf k}) \equiv \delta u(\omega, {\bf k}) - \delta u_L(\omega, {\bf k})$. Evaluating the effective action in \eqref{efaction} and read off the correlation functions
\begin{align}
G^R_{J^tJ^t}(\omega,k) =& - \frac{\sigma k^2 
}{i\omega(1-i\omega \tau_Q) - D_n k^2} - \frac{ n^2 k^2 (1-i\omega\tau_Q)}{(\epsilon+p)((\omega^2 - v_s^2k^2)(1-i\omega\tau_Q)+ i\omega D_\pi^\| k^2)}, \notag \\  
&  + \frac{2 \sigma \omega^2}{|1 - i \omega \tau_Q |^2}\\  G^R_{J^i_\perp J_\perp^j}(\omega,k) &= \bigg( \frac{i\omega\sigma}{1-i\omega\tau_Q} - \frac{n^2 i\omega (1-i\omega \tau_Q)}{i\omega(\epsilon+p)(1-i\omega\tau_Q) - D_\pi^\perp k^2} \bigg) k^{ij}.
\end{align}
which can be matched to fluctuation-dissipation perturbative calculations \cite{kovtun}
\subsection{Relation to the  linear response approach with thermostatic background}

The current $J^\mu$ can generically be parametrized as
\begin{equation}
  J^\mu = J_0^\mu + \mathcal{J}^\mu  
\end{equation}
where $J_0^\mu $ is the bare current and $\mathcal{J}^\mu$ is a mixing between fluctuations and perturbation not necessarily small. 
The geometry conservation implies 
\begin{equation}
   \partial_\nu J^\nu = 0. 
\end{equation}
It is important to note that considering only the exponential associated with $D_{\mu\nu}$, we can rewrite the partition function \eqref{genpfunc} as
\begin{align}
Z &= Z_0  \,    \ave{\exp^{- (J - J_0)}}_0 
\end{align}
where $ Z_0 =  \exp\left[ - C_{\mu \nu,\alpha,\beta}\left(\Sigma,\Sigma'\right) \left(T^{\mu \nu}(\Sigma)-\ave{T^{\mu \nu}(\Sigma')}\right)  \left(T^{\alpha \beta}(\Sigma')-\ave{T^{\alpha\beta}(\Sigma')}\right)\right] $. Instead of \eqref{cfr}, we are not gonna separate the fast and slow modes as \eqref{pqppqp}. It follows that the free energy $F = - \ln Z$ can be expanded as
\begin{align}\label{asd}
F &= \sum_{k=0}^{\infty} F_k  ~~~;~~~
F_0 \equiv - \ln Z_0 \ , \\
F_k &\equiv -  \frac{(-1)^k}{k!} \langle (J - J_0)^k \rangle 
~~~~~~~~(\mbox{for}~k\ge1) \ ,
\end{align}
Potential perturbation $\longleftrightarrow$ Gauging a symmetry $\longleftrightarrow$ Current emerges naturally. The $n-$point Green’s function computed using the variational formulae
\begin{equation}
  \ave{\underbrace{J^{\mu}(t_1,x_1)... J_{\mu}(x_n,t_n)}_{n}}= \frac{\delta^n \lnz}{\delta \left. A_{\mu} \right|_{x_1,t_1}... \delta \left. A_{\mu} \right|_{x_n,t_n} },
  \label{dpartfunc}
\end{equation}
We implement a geometric constraint on the kinematics of the Green function to restore the general covariance 
\begin{equation}\label{cace}
 \bigg\{ \nabla_\mu \mathcal{D} (\Sigma,\Sigma^\prime) -    \frac{\delta (\Sigma^\prime - \Sigma)}{\sqrt{-g}} \bigg( A^\lambda \nabla_\mu \ave{ \partial_\lambda F_{\mu\nu}}_{\Sigma} + A^\mu \nabla_\nu \ave{\partial_\mu F_{\nu\lambda}}_{\Sigma} + A^\nu \nabla_\lambda  \ave{\partial_\nu F_{\lambda\mu}}_{\Sigma} \bigg) \bigg\} = 0
\end{equation}
where as usual $F_{\mu\nu} = \partial_\mu A_\nu - \partial_\nu A_\mu$. We first examine the analytical characteristics of \eqref{zubpart} and since we are able to use the ansatz \eqref{gaussansatz}, the propagator takes the form
\begin{equation}
 \mathcal{D}_{ \mu \nu}=\mathcal{P}_{\mu \nu}\left(\Sigma^\mu\right) f\left(\beta,J_{\mu}, J_{\nu }',\Sigma\right) ,
\end{equation}
where the projector is $ \mathcal{P}_{\mu \nu} \propto 
D_{\mu \nu} $ and the function $f$ has a Gaussian form 
\begin{equation}
\label{gaussianansatz}
f(...) \sim \prod_{\Sigma(x),\Sigma(x')} \exp \left[-\frac{1}{2} \left( J^{\mu} (\Sigma (x) )-\ave{J^\mu(\Sigma (x))}\right) D_{\mu \nu} (\Sigma (x),\Sigma (x')) \left( J^\nu (\Sigma(x'))- \ave{J^{\nu}(\Sigma (x '))}\right) \right],
\end{equation}
To compute the transport coefficient, we perform a perturbative expansion in \eqref{gaussansatz} step-by-step
\begin{align*}
 & \lnz \simeq \left. \lnz \right|_0 - \frac{1}{2!} \textcolor{black}{\int_{\Sigma,\Sigma'}} \left. \frac{\partial^{2} \lnz}{\partial A_\mu \partial A_\nu} \right|_0 \left[   \left( J^{\mu}(\Sigma)-\ave{J^{\mu}(\Sigma')} \right)\left( J^{\nu}(\Sigma) - \ave{J^{\nu}(\Sigma')} \right)  \right] + \dots  
\end{align*}
Since the two currents are not symmetric and obey different directions between $\Sigma$ and $\Sigma^\prime$. Curvature arises from the global geometric structure of the manifold instead of torsion.  
\begin{equation}
D_{\alpha \beta } \left( J^{\alpha} - \ave{J^{\alpha} }\right)\left( J^{\beta}-\ave{ J^{\beta}}\right) \rightarrow D^{\prime}_{\alpha \beta} \times \text{diag} (\lambda_{1,2,3,4} )^\gamma \times \epsilon^{\alpha \beta \gamma \rho } \times \text{diag} (\xi_{1,2,3,4} )^\rho
\end{equation}
where $\epsilon^{\alpha \beta \gamma \rho }$ is the Levi-Civita tensor. Performing a lorentz transformation in the eigenvalues of the current 
\begin{equation}
    J^{\mu} -\ave{J^{\mu}}= \Lambda^\mu_{\;\alpha}\, \text{diag}(\lambda_{1,2,3,4})^{\alpha}  \eqcomma \Lambda^\mu_{\;\nu} \textcolor{black}{=\exp\left[  \int \frac{i}{2}  d\omega_{\alpha \beta} \left( M^{\alpha \beta} \right)^\mu_{\; \nu} \right] },
    \label{lorentzdec}
\end{equation}
where $\Lambda \in O(1,3) $. Let us change the solutions from lab to the comoving frame 
\begin{equation} \lambda_0 \simeq \xi_0 \simeq n_o \eqcomma \lambda_{1-3} \simeq \xi_{1-3} \simeq 0 \eqcomma \Lambda^\mu_\nu u^\nu \simeq (1,\vec{0})
\end{equation}
where $n_o$ is the equilibrium charge density
\begin{equation}
\label{gengauss}
\mathcal{Z} \simeq \int \prod_{\mu \nu} d\omega_{\mu \nu}  \exp \left[ - \omega_{\mu \nu} \omega^{\mu \nu}  D_{\mu }^{ '\mu }\right] \eqcomma D_{\mu \nu}^\prime  =  \left( \Lambda^{\alpha}_{\mu} \right) \left( \Lambda^{\beta}_{\nu} \right) D_{\alpha \beta} ,
\end{equation}
Integrating in both sides to obtain the Green function
\begin{equation}
\int d \Sigma^\prime \nabla_\mu \mathcal{D}^{\mu\nu} (\Sigma,\Sigma^\prime) =  \int   d \Sigma^\prime   \frac{\delta (\Sigma^\prime - \Sigma)}{\sqrt{-g}} \bigg( A^\lambda \nabla_\mu \ave{ \partial_\lambda F_{\mu\nu}}_{\Sigma} + A^\mu \nabla_\nu \ave{\partial_\mu F_{\nu\lambda}}_{\Sigma} + A^\nu \nabla_\lambda  \ave{\partial_\nu F_{\lambda\mu}}_{\Sigma} \bigg)  
\end{equation}
where \( g = \det(g_{\mu\nu}) \). The divergence of an antisymmetric tensor and and the covariant divergence of an antisymmetric tensor can be written as
\begin{equation}
\nabla_\mu F^{\mu\nu} = \frac{\partial_\mu \left( \sqrt{-g} \, F^{\mu\nu} \right)}{\sqrt{-g}}  = \mu_0 J^\nu \ \ \ \nabla_\mu J^\mu = \frac{\partial_\mu \left( \sqrt{-g} \, J^\mu \right)}{\sqrt{-g}} = 0,  
\end{equation}
Using $ \Gamma^\mu_{\mu\lambda} = \partial_\lambda \ln \sqrt{-g}$. Applying in \eqref{cace}, we finally find
\begin{align*}
\mathcal{Q}^\nu = \int d \Sigma \bigg[  \nabla_\mu D^{\mu\nu} (\Sigma,\Sigma^\prime) + \mu_0 \sqrt{-g} \sum_i \frac{1}{\lambda_i \xi_i } A_\mu \nabla_\alpha J_\beta \epsilon^{\mu \alpha \beta \nu} \bigg]  
\end{align*}
where $Q^\nu$ is a constant of motion. The Helmholtz free energy \eqref{asd} can be expanded in a Taylor series. The thermodynamic form is
\begin{equation}
F(N) \approx F(N_0) + \frac{1}{2} \left( \frac{\partial^2 F}{\partial N^2} \right)_{T,V} (\delta N)^2 + \mathcal{O}(\delta N)^3
\end{equation}
and defining $\delta N = N - N_0$. The density fluctuation is
\begin{equation}
\ave{\mathcal{N}^{2}} - \ave{\mathcal{N}}^{2}\equiv k T V \frac{\partial^2 P }{\partial \mu^2} = k_T \Bar{\mathcal{N}} \frac{k T}{v}  \Rightarrow \ave{\Delta J_{\alpha} \Delta J_{\beta}} \sim \left. \frac{\partial^{2} \lnz}{\partial A_\alpha \partial A_\beta} \right|_0   d\Sigma_\alpha d\Sigma_\beta,
\end{equation}
The covariantized version is
\begin{align}
\frac{\partial D^{\prime}_{\mu \nu}}{\partial D_{\mu \nu}} + \frac{\partial \Sigma_\mu \partial \Sigma_\nu}{\partial A_\mu \partial A_\nu} = 1
\end{align}
The transport coefficients in terms of our previous results are given via the usual Kubo formulae
\begin{equation}
\label{transcoeff}
\mu \sim  \lim_{k \rightarrow 0} k^{-1} \text{Im} \tilde{D}_{xt}^{'} \eqcomma \phantom{A}
%c_s^2  \sim \lim_{k \rightarrow 0}  k^{-1} \text{Re} \tilde{D}_{tt}^{'} \phantom{A} \eqcomma
\tau_Q \mu \sim  \lim_{k \rightarrow 0} k^{-2} \text{Im} \frac{\partial^{2}}{\partial k^{2}} \tilde{D}_{xt}^{'}. 
% \phantom{A} C_V \sim \max_k Re \tilde{C}_{xxxx}^{'}, ....,
\end{equation}
This shows how the usual calculations on a thermostatic background match with the non-perturbaive approach developed in this work.  Of course, in a hydrostatic background thermal field theory techniques can be used to calculate ''non-gaussian'' corrections in the form of higher order Feynman diagrams, as was done in \cite{kovtun,delacretaz,xinan} and others.     In such calculations however non-Gaussianities appear as corrections to Gaussian parameters, since they enter into the propagator,

\end{document}